# Iterative Design of Gestures During Elicitation: Understanding the Role of Increased Production

Andreea Danielescu[*]

Accenture Labs, San Francisco, California, USA, andreea.danielescu@accenture.com

David Piorkowski

IBM Research, Yorktown Heights, New York, USA, djp@ibm.com

Previous gesture elicitation studies have found that user proposals are influenced by legacy bias which may inhibit users from proposing gestures that are most appropriate for an interaction. Increasing production during elicitation studies has shown promise moving users beyond legacy gestures. However, variety decreases as more symbols are produced. While several studies have used increased production since its introduction, little research has focused on understanding the effect on the proposed gesture quality, on why variety decreases, and on whether increased production should be limited. In this paper, we present a gesture elicitation study aimed at understanding the impact of increased production. We show that users refine the most promising gestures and that how long it takes to find promising gestures varies by participant. We also show that gestural refinements provide insight into the gestural features that matter for users to assign semantic meaning and discuss implications for training gesture classifiers.

## 1 Introduction

The use of 3D depth cameras to recognize gestures continues to rise, especially now with the push to augmented and virtual reality systems. Starting with the release of the Microsoft Kinect in 2009, more and more 3D depth cameras have made their way onto the marketplace including the Structure Sensor for the iPad, the

---
[*] Research was conducted while at Arizona State University.

iPhone, the Microsoft HoloLens, the Oculus line of products, and the integration of the Intel RealSense camera into smartphones and drones. While plenty of hardware that supports gesture recognition exists, full-body and free-space gesture interfaces are still not commonplace, due in part to the fatiguing and unnatural nature of most gesture-based systems. For example, *dwell*, a gesture which is commonly used to select items on screen, is known to result in "gorilla arms." That's when users experience sore and tired arms due to interactions in which they must hold their arms up in the air with no support. That violates Nielsen et al.'s ergonomic principles [36] and puts a significant amount of strain on the shoulder joint, which is known to fatigue quickly [26]. In addition, the health concerns brought on by the global pandemic have led to a renewed interest in touchless interfaces, putting a stronger focus on free-space gestural interfaces [7, 17, 47].

To design gestures that are intuitive [58], not fatiguing [22], discoverable [3], memorable [2, 35] and learnable [2], gesture elicitation studies have often been used for various types of applications and modalities [1, 5, 6, 11, 14, 27, 28, 31, 32, 34, 39, 41, 43, 58]. Gesture elicitation is a participatory design methodology in which users are presented with referents (an action's effect) and are asked to provide symbols (the interactions that could result in that referent) [56, 58]. Gesture sets produced with this method have been shown to be simpler and more preferred over ones defined by HCI professionals [34].

Originally, gesture elicitation studies asked participants to produce only one symbol per referent (or one one-handed gesture, and one two-handed gesture per referent) [58]. One drawback of this method is that participants may suffer from legacy bias, in which prior experience with existing interactions (i.e. touch, desktop, Microsoft Kinect) influences which gestures users will specify, making it difficult to uncover new or modified gestures that take full advantage of the capabilities of an emerging interaction modality [33, 45].

Several gesture elicitation methodologies have been proposed to address legacy bias. For example, Morris et al. suggested three different methods to move beyond legacy bias: *priming*, *production* and *partners* [32, 33]. Of these, production – in which participants are asked to produce more than one symbol per referent -- has been adopted by various researchers [6, 20, 24, 27–29, 33]. However, the way in which increased production is implemented in each of these studies varies. In Morris et al.'s pilot study, participants were asked to produce as many gestures as they could come up with until they ran out of ideas – what we call *unlimited production* [33]. In this pilot, Morris et al. noticed that, while participants favored the third gesture they produced, participants' gestures also diminished in variety. This has led to most studies inspired by the original Morris et al. study to ask participants to produce a maximum number of 3 symbols per referent [6, 20, 29].

Morris et al., however, left it as an open question to better understanding whether production elicits gestures which are unbiased by prior experience, or whether latter proposals are just irrelevant variations of prior ones, noting: "Consider production. Does it truly increase the variety of proposed interactions, or are downstream proposals from the same participant simply minor variants on earlier ones?" [33]. Existing research has produced contradictory outcomes on the effect of increased production on legacy gestures [20, 33, 55]. We offer a slightly different interpretation: that those minor variations are not irrelevant, and that these latter proposals are refinements of prior gestures which preserve the gestural features that matter most to users. This would suggest that features which are constant across refinements are more semantically salient. Additionally, we believe that a motive for users to refine a preferred gesture is to make it more natural and thereby less fatiguing. Therefore, gesture recognition systems should be designed to discriminate those features which are constant, while remaining invariant to features which are frequently refined.

To explore our hypothesis, we conducted a study in two parts. The first part was a gesture elicitation study for full-body interactions with a public display using unlimited production – increased production that does not impose a limit on how many gestures a participant can propose. Unlimited production enabled participants the freedom to explore gestures and their variations through iteration on preferred gestures in a way that other production methods may not. Although the focus of this study was not on the elicited gesture set per se, the use of the public display context may have influenced the types of gestures produced, such as by encouraging a focus on performative gestures. These considerations are addressed further in the discussion. The second part of the study was a retrospective, where the participant was asked to look back on the gestures they performed and evaluate them. The retrospective was aimed at reducing any ambiguity in the performance of the gesture by allowing researchers to ask participants clarifying questions and to identify gestures which were considered refinements of previous gestures. This enabled us to evaluate the effect of increased production and the exploration of iterative refinements of gestures by participants. This paper, therefore, contributes the following:

- A follow-on unlimited production elicitation study to Morris. et al.'s original pilot that aims to shed light on user's iteration of and refinement of gestures.
- An analysis of gesture refinements showing that 10% of productions are refinements which provide insight into users' mental models by identifying the most relevant features of a gesture.
- A discussion of how these features could be used to design better gesture recognition systems.

## 2 Related Work

### 2.1 Gesture Elicitation

Since its first application to stylus gestures [56], end-user elicitation has been used for a variety of emerging interaction and sensing technologies, such as tabletop multi-touch gesture sets [11, 31, 34, 58], pen-based interactions [14], mobile phone motion gestures [43], drone interaction [5] in-vehicle interaction [28], browser-based interaction [32], TV control [51], full-body interactions with marionettes [16, 27], augmented and virtual reality [39–41], and many others. Beyond hand gestures, researchers have also explored foot gestures [1, 13], a form of hand gesture called micro-gestures [6], and full-body gestures [3, 8].

When first introduced, gesture elicitation studies asked participants to produce only one symbol per referent [56]. In some cases, they were asked to produce one one-handed gesture and one two-handed gesture per referent [58]. With the introduction of partners, priming and production, some studies have incorporated the use of increased *production*, in which participants are asked to produce multiple symbols per referent [6, 20, 24, 27–29, 33]. However, while in Morris et al.'s original pilot, participants were asked to produce as many gestures as they could come up with until they ran out of ideas (unlimited production) [33], many other studies capped the number of gestures they asked participants to produce, usually at a maximum of 3 gestures [6, 20]. A few studies have chosen other variations. In one case, the participants were instructed to produce at least 3 [24] and in another case, participants produced between 2 and 5 [28]. In this study, we use the original production method suggested by Morris et al. [33] to further explore the impact that unlimited production has. This method is most likely to reveal additional refinements that participants may think of but not act out if they were instructed to produce a specific number of unique gestures per referent.

In gesture elicitation studies, it is common to identify a non-conflicting gesture set by aggregating proposed gestures across participants for each referent and calculating an agreement score [52, 53, 56]. Morris [32] proposed the *max-consensus* and *consensus-distinct* metrics to accommodate studies in which an arbitrary number of interactions are proposed per referent. The max-consensus ratio is equivalent to the percentage of participants that suggest the most popular proposed interaction for a referent, while the consensus-distinct metric is the percent of the distinct interactions proposed for a given referent (or referent/modality combination) that achieved a specific consensus threshold among participants. Usually, a default consensus threshold of two is used, which means that at least two participants proposed the same interaction. These two metrics provide a peak and spread of agreement to help guide gesture selection.

Recently, agreement scores have been criticized as misleading as they do not account for chance agreement and they are based on assumptions that may not be true [48]. to correct for these issues, Tsandilas [48] proposes an agreement score calculation based on inter-rater reliability with a chance agreement correction, but this method still does not correct for the significant variability in the features qualitatively coded and the coding process. To address this, Vatavu proposes the use of a dissimilarity-consensus method, which leverages a computed objective measure of consensus between users' gesture preferences [50]. As the focus of this paper is not on identifying a resultant gesture set, we do not use the dissimilarity-consensus method. Instead, we opt to calculate the max-consensus and consensus-distinct scores as an indication of the variety of gestures produced across participants.

## 2.2 Addressing Legacy Bias

Multiple studies have found evidence of legacy bias, with gestures produced by users heavily influenced by mouse interactions [10, 32, 58]. For example, users explicitly mentioned that they would pretend their hand was a mouse they would use to point at objects on screen [32]. Considering the ubiquity of these technologies this is not surprising—novel systems require novel thinking, and most users resort to habits learned in more familiar scenarios to reduce cognitive load.

While legacy bias is thought to produce more fatiguing and less appropriate gestures for novel interaction modalities [33], it may also have some benefits. Culturally shared metaphors, for example, are one reason for legacy bias and shared metaphors may lead to higher agreement scores [58], discoverability and learnability [56]. Gestures based on metaphors have also been shown to be learned more quickly [25]. Köpsel and Bubalo argue that we can benefit from legacy bias, as it can be a helpful tool to gently introduce new forms of interaction, such as gestures or multimodal interfaces, to the general public [23].

However, Cafaro et al. found that legacy bias did not increase discoverability in their study [3]. In fact, these gestures were less discoverable than gestures generated through their framed guessability method, which did not generate as many gestures that were influenced by legacy bias. They posit that this is because by situating gesture elicitation within a "frame" (i.e., scenario), generated gestures are more similar to one another than those generated using other elicitation methods. This begs the question of whether legacy bias has any benefit at all.

Beyond framed guessability, various other methods have been developed to address legacy bias in elicitation studies. One approach uses soft-constraints [44] to penalize users for physical movement of their arms by using wrist weights during the elicitation process. This approach was explicitly aimed at producing fewer fatiguing gestures. Using this approach, Ruiz and Vogel showed that participants produced gestures with subtler arm movements and a wider range of gestures using other parts of the body. A similar approach to elicit gestures that were less fatiguing was used by Siddhpuria et al. [46], in which they explored at-your-side hand gestures. Finally, to calculate fatigue in arm gestures, Hincapié-Ramos et al. introduced the consumed endurance measure and associated workbench, which calculates fatigue from the "gorilla arm" effect based on amount of shoulder joint rotation [18, 19].

Other approaches include use of *priming*, *production* and *partners* introduced by Morris et al. [32, 33]. *Priming* was inspired by findings that participants who performed tasks with physical objects before using touch-tables were less likely to use pointing-only interactions [8, 37]. *Production* was inspired by ideation and brainstorming processes in design [8]. The use of *partners* was also inspired by group brainstorming methods used in design as common practice such as in [9]. While *priming* has seen mixed results [3, 20, 33], the use of *partners* [29] and increased *production* has been adopted by various researchers [6, 24, 27–29]. However, prior work has shown contradictory outcomes on the effect of increased production on legacy gestures. Some studies showed little to change in the legacy gestures participants produced [20], and others resulted in participants producing more novel gestures [33]. Yet other studies found that the effect of production reduced legacy bias in those referents that were most prone to legacy biased proposals [55]. From these results, more research must be conducted to understand the impact of increased production on legacy bias. In this paper, we explore the use of unlimited *production* to explore iterative refinements of gestures and what they tell us about users' mental models. We also aim to provide additional information that may ultimately shed light on the effect that increased production has on legacy gestures.

## 3  Method

For an in-depth exploration into the features that matter to users, we conducted a gesture elicitation study for full-body interactions with a public display with 22 participants, which is consistent with other elicitation studies [54]. We gave participants a scenario and showed them 10 referents (i.e., tasks) related to that scenario, which will be described in more detail below.

### 3.1  Participants

Twenty-two participants over the age of 18 (average age = 27 yrs., S.D. = 6.53) were recruited from the larger university population (12 females, 9 males, 1 non-binary).  All participants were right-handed and had various levels of comfort with technology. Half of the participants had previously used some kind of free-space gesture recognition system before (such as the Microsoft Kinect or Leap Motion) and about half (n = 11) play video games**.**

### 3.2  Environment

The study was conducted in a lab with a large projection display that was approximately 6 ft. (2 m) from the chairs in which participants were seated at the beginning of the study. The surrounding area was large enough for the participant to move around in (it was approximately a 6 ft. x 6 ft. open area). Participants were videotaped by two cameras for the duration of the study. One camera was set up in the front of the space to the right of the projection screen and angled to capture a front view of the participant. The other was set up behind participants, also on the right, to capture the gestures in the context of what users saw on screen and to capture instances in which users turned away from the display. Cameras were positioned to be out of the way of any movement the participant may like to perform. See Figure 1Figure 1 for details of the space.

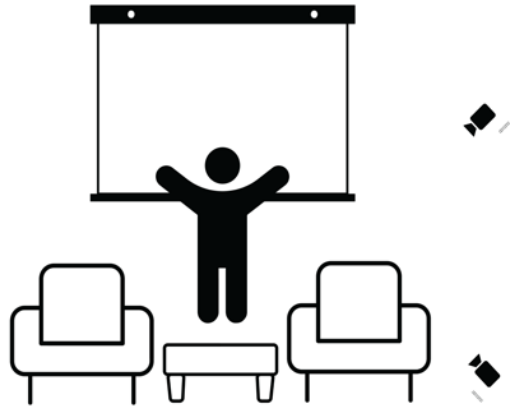

Figure 1. Cameras were placed at 45 degrees in front and behind the participant while they interacted with a projection display in a lab. "Ottoman" icon by Nathan Thomson, "sofa chair" icon by sanjivini, "video" icon by Funky, "person" icon by Hat-Tech, and "projector screen" icon by Random Panda from the Noun Project.

**3.3 Procedure**

At the beginning of the study, participants were asked to fill out a short demographic survey. The remainder of the study was split up into two parts, each of which lasted approximately 45 minutes. Part I was an elicitation study, while part II was a retrospective of the elicitations. Participants were given a short break (5 – 10 minutes) in between, with the study lasting approximately two hours. All participants were compensated for their time with a $20 USD gift card.

*3.3.1 Part I: Gesture Elicitation*

In the first part of the study, we conducted a gesture elicitation study for full-body interactions with a public display. At the beginning of the elicitation section, participants were provided with the following scenario:

> "Pretend you're in a shopping center when you come across an interactive display providing information about national parks. You've always wanted to visit Yellowstone, Glacier, and Grand Teton National Park, so you decide to learn more about what there is to see and do at these locations."

Participants were shown referents (i.e., tasks involving actions to be performed by the participant such as selecting an item on screen) related to the scenario described above. The referents included interacting with photos of animals and landscapes found in national parks. See Figure 2 for an example of one of the referents shown in this study. Consistent with Morris et al.'s original unlimited production pilot, participants were asked to produce as many symbols (i.e., gestures) for each referent as they could come up with. We did not stop the participant from generating any gestures as long as they still had ideas, nor did we set a target number of gestures for them to produce. We also did not give them any instructions on how different or similar gestures should be. We instructed them not to constrain themselves to current technological limitations and to pretend like the display could recognize any movement they made. They were asked to think-aloud during the elicitation part of the study and to start by standing.

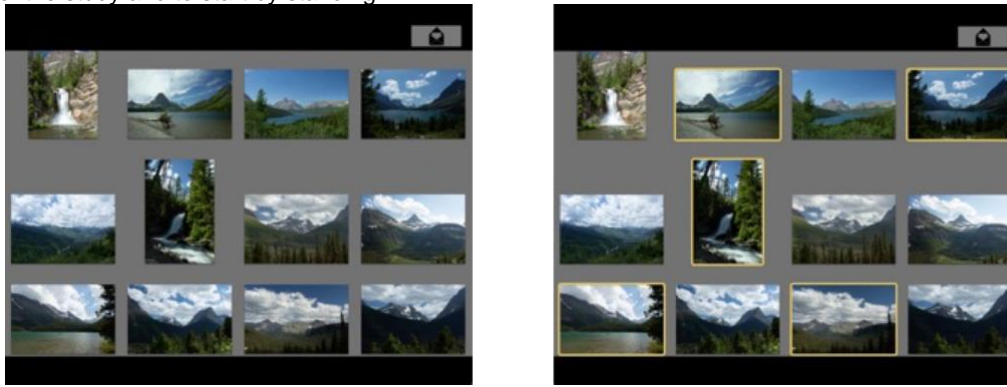

Figure 2. The "multi-object select" referent showed a set of photos arranged in a grid layout (left image) being selected one at a time until multiple images have been selected (right image). Study participants were asked to identify free-space gestures they would use to select the objects.

Once the participant ran out of ideas for gestures to perform for a particular referent, they were asked which of the gestures for that referent was their favorite and least favorite similar to [33]. We defined favorite to mean the gesture which the participant was most willing to perform in public to limit variability that might occur due to interpretation or alternate motivations for picking a favorite. Note that some participants identified multiple favorites or least favorites per referent.

Throughout the elicitation, participants were shown 10 referents. The first referent (*page left*) was used to get people familiar with the task, and therefore was not coded. The order of the remaining nine referents was randomized and each one of the participants saw one of four possible referent orders. The full list of referents that participants were presented were *page right, drag, scroll fast, scroll* (slow), *select, multi-object select, zoom in, zoom out*, and *deselect*.

During the elicitation portion of the study, the experimenter noted a short description of each gesture as it was performed, and any observations or questions they had about the gesture so that follow-up questions could be asked during the retrospective. The experimenters also asked a few clarifying questions during the elicitation phase when necessary, but this was kept to a minimum so that it didn't interrupt the participant's flow.

*3.3.2 Part II: Retrospective*

In the second part of the study, experimenters conducted a retrospective with the participants. The goal of the retrospective was to clarify any ambiguity observed during the elicitation, to gather user ratings of each gesture without inhibiting their creative flow during the elicitation process, and to identify whether participants thought that each produced gesture was independent of the others or an iterative refinement.

During the break, the experimenter copied the recorded videos of the elicitation section onto a Mac Mini and loaded the videos in ELAN[1], making sure they were time aligned. Participants sat in one of the chairs in the space and were shown the videos of themselves performing the gestures. They were asked to rate each gesture on a five-point Likert scale for discoverability, ease-of-use, and appropriateness to the action (see supplemental materials for full question text). The experimenter also asked the participant if a gesture was a refinement of a previous one, and if so which one. In cases in which the experimenter asked any additional clarifying questions that they thought would provide necessary and missing information, user's answers were noted.

---
[1] https://archive.mpi.nl/tla/elan

## 4 Coding and Analysis

### 4.1 Coding Tasks and Analyses

To prepare the gestures for further evaluation, we performed two coding tasks. We first segmented the gestures, identifying the start and end points for each gesture that participants made in the recordings. Second, after we had segmented the gestures, we qualitatively coded each gesture into its component parts to identify which body parts were involved, what direction they moved and the timing and speed of these movements. For both these coding tasks two researchers followed a consensus-coding approach [30], where coding guidelines were developed iteratively and disagreements between the researchers were discussed and resolved during the coding process.

### 4.2 Gesture Segmentation

For segmentation, we started by defining the temporal boundaries of a gesture. Any movement related to the user preparing for the gesture was ignored unless the user indicated in the think-aloud that this movement was important. Many participants repeated the gestures several times during the elicitation process, so only the first time the gesture was performed was segmented. We chose the first gesture because they tended to be crisper than the repeat gestures that followed, thus easing the coding process. During the coding process, additional repeats were checked for inconsistencies and noted (e.g., if users switched from a 1-finger to 2-finger point). Segments of the video that provided additional information via think-aloud or that contained movement in which the participant was exploring the gesture space were also annotated for reference.

We segmented a total of 1117 gestures (1442 total annotations, including exploratory gestures, additional information from the think-aloud, and complementary gestures – for example, a gesture that was meant to *deselect* when the participant was shown a *select* referent – that didn't match the referent) across 18 of the 22 participants. Four participants were excluded due to either missing elicitation videos or technical recording errors.

### 4.3 Gesture Coding

After the gestures were segmented, two of the researchers qualitatively coded the gestures. Due to the large number of gestures (1117), we selected 10 of the 18 participants' data for further coding. We intentionally selected participants to maintain an even distribution of video order, participant gender (half were female), and experimenter. Consequently, we coded 545 gestures. Although there is a possibility that the 8 excluded participants performed differently than the 10 we included, additional analysis showed that we achieved data saturation with the 10 we had [8, 15, 21, 49]. In particular, we calculated the cumulative total gestures each participant added across a random selection of participants. We found that after running 3 participants, 74% of the gestures were already specified. By 6 participants 95% had been specified, and no new gestures were specified after 8 participants. Subsample permutations conducted on n >= 5 also show that running and analyzing 6 participants is enough to identify the most common gestures (those with the highest consensus scores) across participants, we settled on 10 participants to identify long-tail gestures with lower agreement scores that are less likely to be influenced by legacy bias. Considering that this is a formative study, we view this as an acceptable risk.

To properly understand how gestures are refined and iterated upon, we chose to code gestures at a finer granularity than other studies. As with similar studies, we began with coding for high-level gestures (e.g., "swipe, hover, point"), side of the body used, an indication of the type or direction of motion, and similar to elicitation studies which focus specifically on hand gestures, the hand configuration (e.g., thumbs up, 1-finger extended, fist, flat hand, etc.). Hand configuration was coded even though evidence from Wobbrock et al. [58] and Morris et al. [34] shows that the number of fingers used is unimportant in touch interactions, because for free-space interactions distinguishing between someone using a 1-finger point, pointing with a flat hand, or pointing with a fist might still be important. Since this study focuses on full-body interactions, which includes gestures such as kicking, jumping, and walking around, we coded all body parts used for each gesture. Full-body interactions often consisted of a combination of gestures, so we also coded the gestures and their associations based on how they were performed (e.g., in sequence, simultaneously, etc.). For example, a sequence may include walking to a specific point in space and then jumping. A simultaneous gesture might include swiping with one arm up and one arm down at the same time. We then split composite gestures (e.g., sequences) into what we called gesture primitives, with a total of 660 gesture primitives across 545 gestures.

We also included features to identify the path more accurately in space that body parts moved through during the users' gestures. These features were inspired by how animation software encodes animations, allowing animators to define the key frames of the movement and interpolating the rest. Likewise, the idea of key frames guided which parts of a gesture's motion to code and which parts of the motion can be trivially interpolated using the information from the gesture's sequence immediately before and after the movement. We, therefore, encoded the beginning and end points of commonly used limbs and the starting stance in cases where it was relevant.

We also coded for the primary joint of rotation to explore the impact fatigue may have had in refinements. The primary joint of rotation was coded as studies show that larger joints (e.g. the shoulder) will fatigue faster than smaller joints (e.g. elbow or wrist) [18, 26]. In cases where multiple joints were involved, we coded the joint that is likely to fatigue fastest.

A total of 49 gesture primitives were identified during the coding process. Example primitives include pinch (any movement where multiple body parts move towards one another after being separated) and shake (short quick movements back and forth of a joint). Once all the gestures were pairwise coded, the experimenters went through all the coded gestures together, and resolved any remaining discrepancies. Some of the gesture codes were further refined. For example, similar gestures that were coded as discrete primitive sequences (like touch and tap) were collapsed into one higher-level gesture primitive (e.g., tap). In another example refinement, a sequence of gestures was collapsed into a single gesture primitive with additional information captured into a different feature (such as collapsing all drawing gestures into one gesture primitive with the path details containing the thing that was drawn, e.g., an "X"). Generally speaking, such refactoring was focused on making the gesture coding more consistent and where possible, smaller without loss of detail. 38 gesture primitives remained after consolidation. For more details on the qualitative coding, see the codebook, sample coding and gesture primitive definitions in the supplemental materials.

# 5 Results

## 5.1 Elicitation Study Methodology Modifications for Legacy Bias

As prior work has produced such contradicting outcomes on the effect of increased production on legacy gestures [20, 33, 55], we take a closer look at the effects of unlimited production on user preference and on legacy bias in this study. We start with user preference.

Given the ability to produce as many gestures as they wanted, participants produced more gestures than many prior studies which generally limited participants to only 3 gestures. They also often preferred later gestures as their favorites. Participants produced between 2 and 10 gestures per referent with an average number of gestures per referent of 5 (SD = 1.74). Within study participants, we observed no difference between the number of proposed gestures from the first to the last referent. To start, we look at the median position of participants' favorite gesture, which was 3 (SD = 2.07) and the median position of the least favorite gesture, which was 4 (SD = 1.68).

Because averages and median calculations can be misleading, we chose to do additional analysis on the favorites and least favorites to better understand how participant preference may be impacted by production and by individual preferences. While 20% of participants' favorites were at position 3, there were even more favorites (30%) at position 1. These 1st position favorites were counterbalanced by favorites that were identified after position 3 though with 33 out of 97 (34.02%) of participants' favorite gestures occurred after three other gestures had been performed. Similarly, 31 out of 89 (34.83%) of participants' least favorite gestures occurred in the first three gestures performed. For a high-level overview of the cumulative total of favorites and least favorite gestures across all participants and referents by gesture number, see Figure 3. Together, these findings suggest that some participants needed an opportunity to experiment with gestures before encountering their preferred gesture.

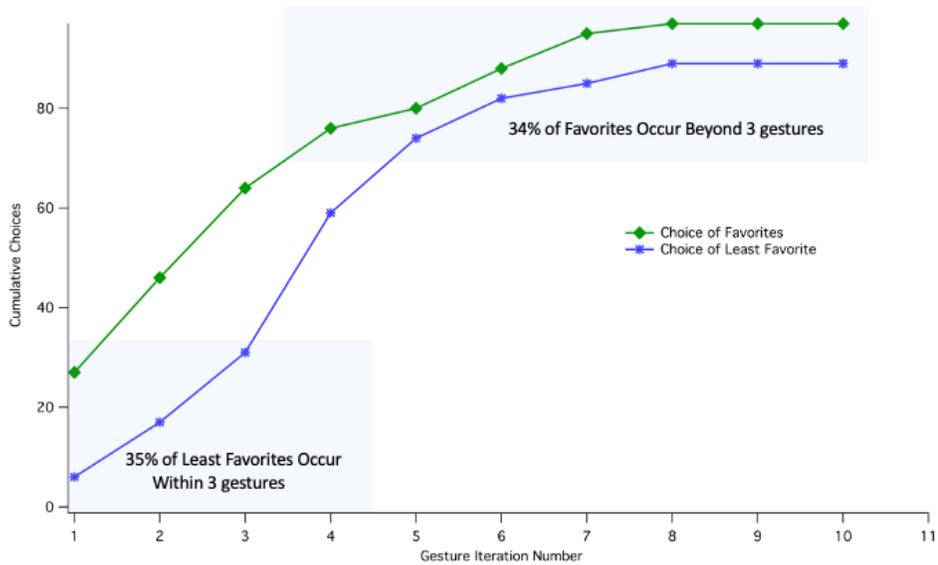

Figure 3: Plot of cumulative frequency of favorite and least favorite gestures across all participants by gesture number. The gesture number indicates the order in which the gesture was produced during elicitation.

Looking at Figure 4, we see individual participant differences in how quickly they settled on their favorite gestures. A couple of participants, such as P3 and P9, skewed heavily towards preferring their first gesture. A majority preferred gestures produced between position 2 and 4. Finally, a few participants (P2, P6 and P8) were just as likely to prefer gestures in positions 6 and 7 as they were to prefer the first or second gestures. Limiting participants to 3 gestures per referent or less as in some prior work [6, 20] without encouraging experimentation, may leave such participants unable to find their preferred gestures, and without the opportunity to consider gestures not informed by legacy bias.

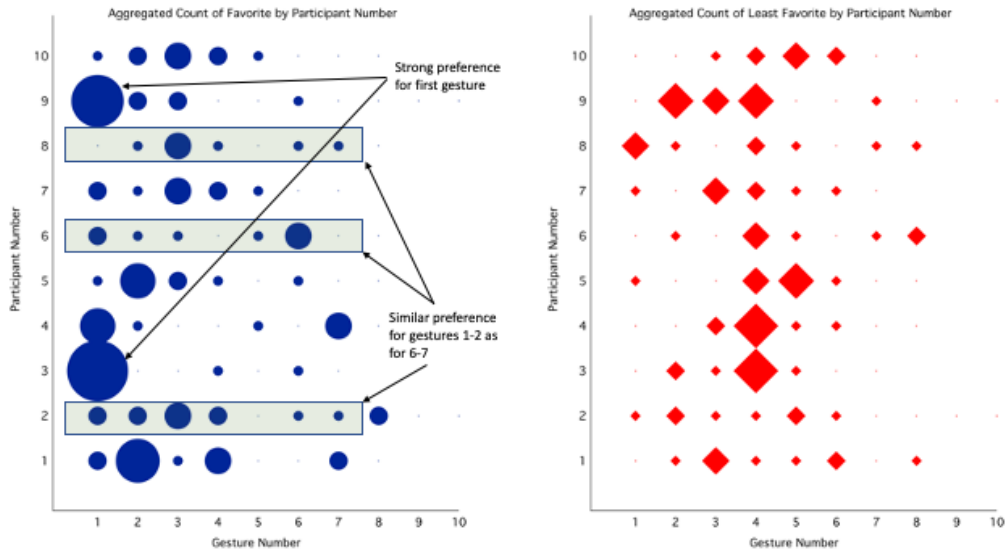

Figure 4: Favorite and least favorite gestures aggregated by participants. The gesture number indicates the production sequence of the gesture during elicitation with empty spaces indicating gestures the participant did not produce at all. Each column shows the frequency of gestures a participant marked as favorite or least favorite at that location, with the size of the dot corresponding to the frequency.

To explore whether user preferred gestures were also ones that were qualitatively identified as the least fatiguing and easiest to perform, most appropriate for the action and having the most guessability, we look at the relationship between these ratings and whether a gesture was a favorite, least favorite, or neither (no preference). We found a strong relationship between the gestures that participants considered their favorite and the ease-of-use, appropriateness and guessability measures. Favorite gestures scored highest, and least favorite gestures scored the lowest (Table 1). An Aligned Rank Transform (ART) ANOVA for nonparametric data [57] shows that the three groups are statistically significant for all three user ratings (guessability ($F(2, 453)$ = 27.44, $p < .001$), ease-of-use ($F(2, 453)$ = 29.21, $p < .001$), and appropriateness ($F(2, 453)$ = 34.35, $p < .001$)). Post hoc comparisons using the Tukey test confirmed that the mean score for all three conditions were significantly different ($p < .001$) for guessability, ease of use, and appropriateness (see Table 1 for means and standard deviations).

Table 1: Means and standard deviations for guessability, ease-of-use and appropriateness ratings across user preference.

| User Preference | Mean Guess. | St. Dev. Guess. | Mean Ease | St. Dev. Ease | Mean Approp. | St. Dev. Approp. |
|---|---|---|---|---|---|---|
| Fav. (n = 95) | 4.08 | 1.16 | 4.62 | 0.80 | 4.55 | 0.77 |
| Least Fav. (n = 87) | 2.67 | 1.39 | 3.37 | 1.35 | 3.08 | 1.36 |
| No Pref. (n = 274) | 3.57 | 1.29 | 4.12 | 1.07 | 3.93 | 1.22 |

Finally, we look at the effect of production on the variety of gestures performed. An effective gesture is one that a broad range of users are likely to intuit, and one that most uniquely represents its referent. Without a breadth of consensus, users would struggle to discover the gestures needed to accomplish the actions they want to take. If an action is well represented by multiple gestures, then participants may have to guess at multiple gestures before finding the right one. These two scenarios are measured using max-consensus (peak agreement for a particular gesture) and consensus-distinct (spread of possible gestures mentioned by two or more participants). Overall, many of the referents had high max-consensus scores while consensus-distinct scores varied. Table 2 shows the max-consensus and the consensus-distinct scores for each referent. Of these, the highest consensus gesture (move arm or hand) for the drag referent is one of the most intuitive as most of the participants came up with the same gesture (90% max-consensus) and there were few other gestures that participants came up with (0.882 consensus-distinct). Conversely, the deselect referent resulted in a gesture with the lowest max-consensus score (60%) and a fairly low consensus-distinct score (0.400), a combination likely to give users a hard time. While we see that some gestures are likely to be guessed by nearly everyone, the unlimited production also allowed participants to propose a variety of gestures that mapped to each referent, suggesting there might be alternative gestures that may be nearly as good as the ones with the highest max-consensus.

Taken together the results in this section show that the gestures users prefer can come anywhere in the unlimited production process (though the majority fall around the third production based on current data). Importantly, we also find that unlimited production may be used to find alternative gestures that may be nearly as good as ones with the highest max-consensus scores. Next, we look at refinements and what they tell us about the features that matter to users.

Table 2: Max-consensus and consensus-distinct scores for each referent. * Move in this case refers to moving one's hand over to the email icon shown on screen to drag the item over. Sometimes, this motion was preceded by a "grab" type movement, but not always.

| Referent | Gesture with Highest Consensus | Max-Consensus | Consensus-distinct |
|---|---|---|---|
| Page Right | Swipe | 100% | 0.421 |
| Drag | Move (arm/ hand) * | 90% | 0.882 |
| Scroll Fast | Swipe | 90% | 0.706 |
| Select | Point | 90% | 0.667 |
| Multi Object Select | Point | 90% | 0.579 |
| Zoom In | Expand | 80% | 0.684 |
| Zoom Out | Pinch | 80% | 0.500 |
| Scroll | Swipe/Slide | 60% | 0.600 |
| Deselect | Swipe/Tap | 60% | 0.400 |

## 5.2 A Closer Look at the Features That Matter

Prior elicitation studies have already shown that users are influenced by existing technologies [10, 32, 58], which means that legacy bias plays role in the types of gestures produce. For example, participants may be predisposed to making gestures that are unnatural, larger and more fatiguing due to limitations of prior technologies that they have used. Absent this bias, participants should produce gestures that are more natural, and more refined than the gestures used with existing technologies, but not necessarily drastically different than gestures currently informed by legacy bias. In this section, we focus on participants' gesture refinements to address our questions regarding which features are most salient to users and briefly explore the role that fatigue plays in users' choice to refine gestures.

We first look at what kinds of gestures are refined and how they compare to gestures that are not, finding that gestures that are refined are those that participants consider most promising (i.e., rate better on guessability, ease of use and appropriateness than the average gesture). We then look at which features change between the originally proposed gesture and the refinements, to understand which features actually matter to participants and which gestures they would consider identical or not. Within these gestures, we take a closer look at how fatigue plays a role in the way that participants refined gestures by looking at the joint of rotation, as research shows some joints (e.g. shoulder) fatigue faster than others (e.g. elbows or wrist) [26] and quantitative gestures of fatigue are based on joint rotations [18, 19].

Out of the 545 gestures coded across our 10 participants, 56 of them were refinements (10.3%). We began by evaluating whether the ratings change between the refinements and the original gestures being refined. Table 3 shows the average usability, ease-of-use and appropriateness ratings for both the refinements and the original gestures. While the average values for the refinements are slightly lower and the standard deviations slightly higher than the originals they refined, an ART ANOVA for nonparametric data [57] showed no significant difference for appropriateness ($F(1, 84) = 0.19$, $p = 0.667$), ease-of-use ($F(1, 84) = 1.14$, $p = 0.289$), and guessability ($F(1, 84) = 2.64$, $p = 0.108$), which indicates that this is either natural variability or that a larger sample size is needed.

Table 3: Means and standard deviations for the guessability, ease-of-use and appropriateness ratings for refinement gestures, the gestures being refined and for the entire set of elicited gestures as a comparison.

|  | Mean Guess. | St. Dev. Guess. | Mean Ease | St. Dev. Ease | Mean Approp. | St. Dev. Approp. |
|---|---|---|---|---|---|---|
| Refinement | 3.67 | 1.27 | 4.13 | 1.02 | 4.11 | 1.16 |
| Original | 4.10 | 1.08 | 4.40 | 0.74 | 4.30 | 0.85 |
| All | 3.87 | 1.20 | 4.26 | 0.91 | 4.20 | 1.03 |

Additionally, of the combined refinements and originals, 29 out of those 114 (refinements + originals) were listed as favorite gestures (25.44%), whereas only 16.90% of all gestures are listed as favorites. Overwhelmingly, participants seemed to refine gestures for which they already had a preference rather than wasting time refining gestures that lacked potential as measured by guessability, ease of use or appropriateness. A two-sample two-tailed t-test indicates this difference is statistically significant ($t(554) = 2.7$, $p < 0.01$).

To address our hypothesis on which features matter most, we analyzed a subset of the features that changed between the refinement and original gesture. The frequency of each feature changing can be found in Table 4. We found that participants primarily refined arm gestures, and nearly all the refinements happened immediately after the gesture they were refining. The most likely features to change between the refinement and original gesture were the palm direction (73.68%) and hand configuration (59.65%) (see Figure 5 for an example). This is not surprising as participants overwhelmingly preferred to use their arms and hands when specifying gestures, comprising 68% of gestures elicited in this study (see Table 5 for a breakdown of the body parts used across all productions by all participants for all referents). This is also consistent with the literature review of gesture elicitation studies conducted by Villarreal-Narvaez et al. [54]. Yet despite the ability to make as many gestures as they liked, it also means that participants did not necessarily take the opportunity to explore less common gestures that could be performed with other parts of their body. Since the majority of existing gesture recognition systems rely heavily on large arm gestures, this tendency may reflect the legacy bias of participants.

Participants also varied gesture direction but did not have any discernable patterns in the change. Gesture path, however, often changed from a straight path to an arching or circular path. The number of gesture primitives changed as well. In some cases, users added extra primitives, and other times simplified. Adding primitives happened in cases where the gesture's intent became more specific. For the side of the body, in all but 1 of the 6 cases, the change was between using both sides of the body or only one side.

Table 4: Frequency of each feature changing across refinements.

| Feature | n (%) |
| --- | --- |
| Palm Direction | 42 (73.68%) |
| Hand Configuration | 34 (59.65%) |
| Gesture Direction | 18 (31.58%) |
| Point of Rotation | 16 (28.07%) |
| No. of Gesture Primitives | 14 (24.56%) |
| Body Part | 13 (22.81%) |
| Gesture Path | 12 (21.05%) |
| Side of the Body | 6 (10.53%) |

Table 5: Percentage of gesture primitives for each part of the body.

| Body part used | % using part |
| --- | --- |
| Arms (hands, elbows, fingers, etc.) | 68.33% |
| Legs (feet, toes, knees, etc.) | 9.70% |
| Voice | 8.33% |
| Gaze | 5.30% |
| Full body | 4.55% |
| Head | 3.33% |
| Brain Machine Interfaces | .45% |

The features that participants change during refinement indicate features that participants do not consider a salient part of the gesture. Correspondingly, the features that do not change, or change little, indicate the features that the participants found most salient, and which would therefore be most useful for gesture recognition systems to identify. We found that many of these feature changes are a consequence of simplifying gestures as shown in the example, leading to our exploration of the impact of fatigue.

To explore whether fatigue plays a role in refinement, we evaluated the primary joint of rotation. In this study, some participants mentioned that the level of fatigue a gesture causes is an important consideration during the elicitation process and for this reason participants wanted to specify gestures that were "more intimate" (P3) – i.e., closer to the body, smaller -- or that "take very little effort" and are inconspicuous (P7).

Participants' choices in refinements agree. Out of the 30 cases in which either body part or point of rotation changed, 10 of those cases are instances in which both changed. In all but 5 of the 30 cases, the movement became smaller (e.g., using the hand instead of the entire arm or changing the point of rotation from the shoulder to the elbow or even the wrist). In 2 cases, the user changed the gesture to use a different finger (e.g., thumb instead of pointer finger). Only in in 3 cases did the movement become larger.

These behaviors suggest that fatigue may play a role in the gestures that participants refine and may be a driving force of the minimization behavior described above. We dig into the implications of these findings in the following section.

## 6 Discussion

### 6.1 The Impact of Unlimited Production on Elicitation

In this study, we explored the effect of unlimited production on user preference of gestures and found that production made a difference in user preference. We see this through the combination of two results: 1) users needed to try a few times before finding gestures they liked as shown by the distribution of favorite gestures across all productions, and 2) these favorite gestures had significantly higher appropriateness, guessability, and ease-of-use scores than other gestures. These results confirm those found by Morris et al. when they introduced the use of increased production [33] and reinforce the value of unlimited production.

These results also seem to contradict the finding that users preferred their first gesture to others by Hoff et al. [20], as our results show that this is true only 30% of the time. Differences in the ways the studies were conducted may account for these contradictory results:

- Our study and the study by Morris et al. told participants to produce as many gestures as they could, while in [20] each participant was instructed to only produce 3 gestures per referent.
- The referents in this study were image-based, while the referents in Hoff et al. [20] were largely text-based and based on a music playlist.

In the case of the production goals, in Hoff et al.'s study [20], participants may have focused too much in producing exactly three gestures as a goal, and therefore conducted some sort of pre-filtering before suggesting any. In our study, we explicitly chose not to impose any limit on production as there is no consensus in the community on what a reasonable limit is for gesture production and unlimited production draws inspiration from research on design outcomes that show there are benefits to brainstorming and iteration [33]. Future work may wish to directly compare a limited and unlimited production method to identify exactly how it changes the gestures elicited from users. For the referents, early research shows that text, such as that presented by Hoff et al. [20] bring to mind different interaction metaphors than the images [8], or that simply the size of the image or the type of menu users were interacting with had an effect [8, 40]. Additional research is required to understand if and how much these factors may have impacted the production process and future elicitation studies must be aware of these possible influences so as not to overgeneralize their results.

The impact of unlimited production on the types of gestures and the variety of gestures produced is nuanced. To understand this effect, we look at the max-consensus and the consensus-distinct scores for the referents. While max-consensus was high for many of the referents (7 of the 9 had max-consensus scores over 80%), consensus-distinct scores varied much more. This means that while nearly all participants specified the same discoverable gesture for each referent, there was also a broad range of additional gestures proposed by participants for each referent. The high max-consensus could indicate that legacy bias still plays a significant role. Further evidence of this is provided by the participants (P3, P13, P15) who explicitly identified that they were drawing from interactions with smartphones. The variability in the consensus-distinct score could partly be the outcome of increasing production as participants did each come up with many gesture options for each referent that were agreed upon by at least 2 of our participants.

One point worth mentioning when it comes to comparing the effect of production on legacy bias is that each study has defined or determined whether a gesture is influenced by legacy bias differently. In their study, Hoff et al. noted that the number of legacy biased gestures did not change over the course of the productions [20]. However, they categorized gestures as being influenced by legacy bias only if participants verbally mentioned they were influenced by prior technology during the study. In comparison, Williams et al. [55] categorized a proposed gesture as a legacy or non-legacy gesture through a consensus of independent votes of three expert raters if the gesture class could be identified as being used with a known device in common usage. In their study, they found that increased production reduced consensus on proposals and was effective in reducing legacy bias by about 12% for those referents that were most likely to be influenced by bias in the first place. Our study also used a consensus approach to determining biased gestures but found that taking a nuanced approach to thinking about legacy biased gestures is more effective. That is to say that it's not just about the gestures themselves, but the features that may make them a poor fit for the current interaction modality. As an example, a 1-finger or one-handed swipe is generally considered a legacy biased gesture, but this gesture can be performed with a lot of structure by the participant (palm facing inward, straight path, discrete start and stop) and with a large motion, e.g., from the far right to the far left of the body, or it can be refined to be more natural. This takes us to the second modification to the methodology, which is the addition of a retrospective.

The addition of the retrospective in the second half of the study allowed the experimenter to explicitly ask participants whether each gesture was a refinement of a previous one, and if so which one. This resulted in our ability to identify that about 10% of gestures were not new gestures suggested by participants but refinements in which users were iterating on gestures that seemed promising. One example of this can be seen in Figure 5, where P9 was suggesting gestures for scrolling through a collection of images. In this example, we can see the distance between the start and end of the gesture decreasing and the palm direction and trajectory of the swipe changing. This refinement in our methodology in combination with the fine-grained qualitative coding that was conducted allowed us to identify features of the gestures that do and do not seem to matter to users. We go into more details about these features in the next section.

Future gesture elicitation studies should strongly consider using unlimited production, where participants are encouraged to come up with as many gestures as they can think of. This has the clear benefit that participants can refine their gestures and come up with ones that are more natural and a better fit for the referents. Researchers should also consider adding a retrospective section to the study methodology, which provides researchers with the opportunity to ask clarifying questions without interrupting participants' flow during elicitation portion. During the retrospective, researchers can better assess which gestures users consider the same, rather than relying on the qualitative coding process alone to do that. A lack of a retrospective section in an unlimited production study would increase the complexity of coding the gestures. Finally, we advocate for future studies to also use a fine-grained approach for identifying gestures along with participant feedback to not only understand what features of the gesture matter to the participant, but to also to identify gaps in current gesture recognition technologies.

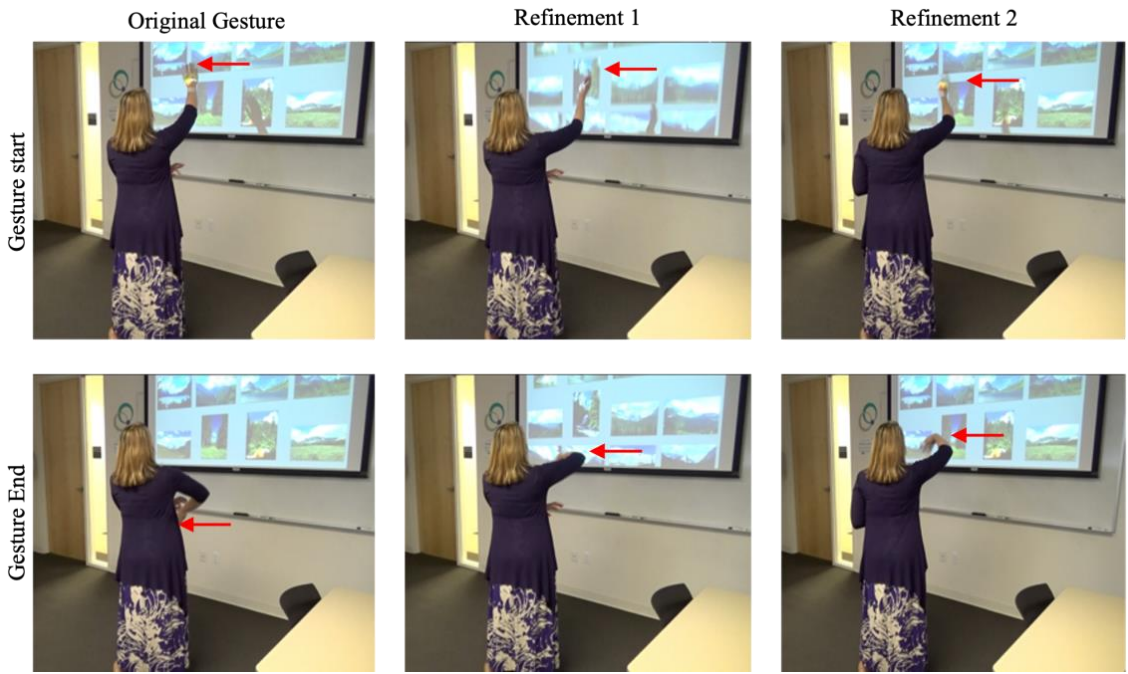

Figure 5: Gesture refinements from P9 for scroll. Red arrows denote the start and end location of the gesture. Note how the gesture becomes smaller (movement between start and end is shorter). Note also that the palm position changes from facing towards the screen to facing downward.

## 6.2 What Refinements Tell Us About User Preference

In our analysis, we demonstrated that users often refined gestures that they already preferred, with gestures that were refined containing a much higher percentage of favorites (25.67%) than the full set of gestures elicited from participants (16.90%). Refinements usually happened immediately, so as soon as participants found a gesture they might like, they put the effort into exploring refinements. Nearly all refinements were focused on arm or hand gestures, not on gestures that were full-body or leg gestures. One possibility for this is that arm and hand gestures are more likely to suffer from legacy bias, as touch and desktop interactions use exclusively hand movements and many Kinect interactions also used arm gestures for navigation and selection tasks. Another reason refinements might focus almost exclusively on arm gestures is that arms inherently have more degrees of freedom than other parts of the body and that so many of the elicited gestures are arm and hand gestures to begin with. Comparing arms to a torso, for example, makes clear the difference in the number of possibilities to be explored.

These results underscore the importance of refinement and iteration as a window into participant's communicative intent and their expectations about the technology they are interacting with. Increased production enabled participants to freely refine gestures, experimenting with more or less natural variants or gestures that were closer or further from legacy gestures – and thus potentially trading off fatigue for discoverability or well-known visual metaphors. Importantly, our results show that refined gestures do not uniformly improve upon original gestures, indicating that refinement is better thought of as an exploratory process in which the results may be more or less favorable. Without the use of a retrospective, it would be extremely challenging to decipher this exploration and make sense of the participant's intentions. Whether through adopting similar methods, or by otherwise encouraging participants to refine and elaborate on gestures, future work should seek to understand when and why refinements occur.

Our results also indicated that some specific features -- such as palm direction and hand configuration -- contain a large amount of variability and are not distinguishing characteristics for users. P9 clearly stated this during the elicitation process for the scroll gesture shown in Figure 5. The finding that hand configuration is not important to users' mental models while gesturing to a public display seems analogous to findings that users do not place importance on the numbers of fingers used in touch interaction [34, 58]. Similarly, hand configuration was only meaningful when the hand configuration itself was the gesture (for example, a thumbs up gesture, or an L-shape). Straight and arching paths were seen to be used interchangeably by users. For example, one participant may swipe from the top left to the bottom right in an arc motion the first time, and swipe from left to right across the body using a straight path a second time. In another case, a user may draw an "X" from the top right the first time around and from the top left the second time around. What matters most for user's mental models in these cases is the resultant shape, similar to how one would write an "X" on paper. These findings have direct implications for gesture recognition algorithms, the variety of the data they're trained on, and the way the data is labeled in supervised learning approaches, as we will discuss below.

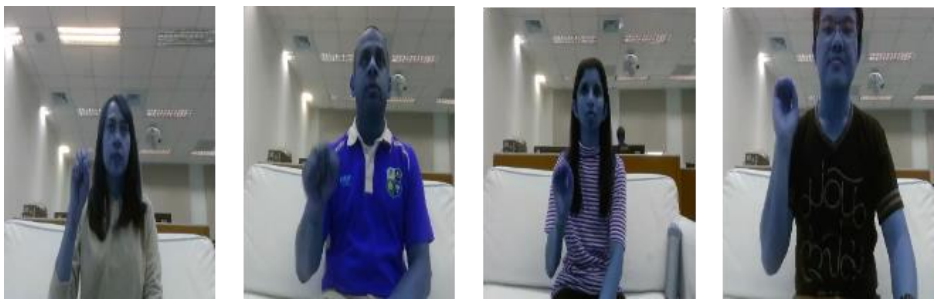

Figure 6: Different images from a data set used to train a gesture recognizer for TV control interaction from [18]. IC4You Dataset Images Copyright Timothy K. Shih and MINE Lab, National Central University. Used with permission.

### 6.3    Implications for Gesture Recognizers

Based on our findings, there are three recommendations developers of gesture recognition systems should keep in mind:

First, real-world variability of gestures is much higher than the variability found in most pre-recorded gesture data sets. For example, Figure 6 shows four examples of a push gesture performed by different people in a controlled manner for a TV control application. This level of precision is not something that people tend to have in normal interactions with technology. Thus, if pre-recorded gestures are needed for, say, training machine learning components of a gesture recognition system, then it would be advantageous to carefully consider how the gestures in question vary when iterated upon by real-world users.

The second is that when conducting any sort of learning through supervised methods, the labels that are used for gestures should match user's mental models and line up with what users believe are similar or different gestures. Features that are not important to participants will have high variability across productions, and the data set and associated labels should mirror this fact. In particular, for public displays, gesture path (e.g., straight vs arching), palm direction, and hand configuration (e.g., 1-finger extended vs. 2-finger extended) should not heavily influence the resultant classification. Ideally, the gesture labels should specifically capture multiple dimensions, such as primary action and second order intentional or unintentional variations influenced by factors such as the location, size, number and type of object users are interacting with on screen, so that the gesture recognition systems can be taught to recognize finer-grained distinctions in these gestures.

The third is that to account for the full variability of natural gestures that are semantically similar for users and which they might want to use, designers should consider a many-to-one mapping of gestures to command. Erazo et. al similarly argued that designers should support many-to-one mappings instead of one-to-one mappings to account for real-world variability [12, 38, 42]. However, our works supports a stronger view that gesture recognizers could leverage many-to-one mappings of gestures that align more closely to users' perceptions of what makes gestures equivalent rather than arbitrary classes of unrelated but interchangeable gestures. Our study's findings on refinements supports this suggestion as some of these refinements would likely be misidentified by current gesture recognizers. These three suggestions may create better, more robust, more usable gesture recognition systems that will lead to higher adoption rates.

### 6.4 Additional Limitations

Like with many elicitation studies, by grounding our study in a particular scenario with specific images, and a UI that wasn't iterated upon, we risked influencing participants in the types of gestures they performed. The size, number, type, layout of the objects, and animations likely influenced certain gestures over others, as previous research has indicated [8, 40]. Future work should seek to better understand the impact of these features on the gestures elicited (e.g., by comparing abstract boxes with images while retaining the same layout, framing, speed, size, etc.) as this will inform the generalizability of gestures elicited in specific scenarios. Furthermore, the nature of the task along with the space in which the study was conducted may have limited participants ability to project themselves in the scenario that we proposed, thus altering the types of gestures performed.

Although the focus of this study was not on the elicited gesture set per se, the use of the public display context may have influenced the types of gestures produced, such as by encouraging a focus on performative gestures. However, we chose to focus on public displays because, while current interactions in public spaces are short lived, we do not anticipate that will always be the case. Due to the performative nature of public spaces, the large awkward gestures of today's gestural interaction systems serve as a deterrent for users to interact with them. The same smaller, more natural gestures that would reduce fatigue, therefore, may also make full body or free space gestural interaction more appealing in public spaces. Similarly, we anticipate that people interacting with public gestural interaction systems will perform a series of gestures to accomplish a task instead of one gesture at a time. We see our work as formative in this regard and more research needs to be done to shed light on how participants intuit gestures for completing a set of tasks instead of specific one-off actions – possibly by leveraging an approach similar to the framed guessabiilty methodology proposed by Cafaro et al. [3, 4].

Our study was limited to 10 participants, and although many of our findings were statistically significant and saturation calculations show that we likely discovered the majority of popular gesture interactions, we caution the reader from overgeneralizing our results. Differences between participants were apparent, as shown in Figure 4, and we anticipate that additional studies will uncover other differences in how gestures are performed and refined, especially when it comes to long-tail gestures.

Furthermore, there is a possibility that the finding that users preferred the third gesture could have been caused by the amount of thought and effort put into creating the gestures, instead of anything about the gesture itself. There are multiple ways that future work could explore this possibility. In our study, we explicitly asked participants about their favorites and least favorites during the elicitation phase so they could remember how it felt to perform the gestures. Future studies could simply ask the participants which gestures were their favorite and least favorite again during the retrospective phase. An additional possibility is to re-order the gestures the participant performed when rating them and re-asking about the favorite and least favorite during the retrospective. Finally, this could be taken even further by conducting the retrospective a few days after the study to verify their choices. We leave such confirmation for future work.

Finally, in our results we saw indications that fatigue may play a role in refinement of gestures, but additional evaluation is needed to understand the extent that fatigue impacted gesture production. Future studies should focus on this specific question more thoroughly, potentially including quantitative measures of fatigue in addition to qualitative assessments by participants.

## 7  Conclusion

In this paper, we presented a gesture elicitation study that used unlimited production, and which incorporated a retrospective during which experimenters were able to ask participants to identify iterative refinements of their gestures. By exploring the impact of unlimited production, we found that limiting production may unnecessarily prevent participants' from finding their preferred gestures as 34% of participants' favorites occurred after the first 3 gestures. Furthermore, we found that about 10% of gestures proposed by users are actually refinements of previously proposed gestures. We show that these refinements can provide valuable insight into users' mental models by identifying those features that are salient to users (i.e., the features that are invariant across iterations) and those that do not matter to them, such as hand configuration and palm direction, that change across iterations. We also saw evidence of gestures getting smaller and using body parts that fatigue less easily (e.g., wrist movements instead of shoulder movements), but the extent to which fatigue impacted gesture production was left for future work. Finally, we offer several recommendations for designers of gesture recognition systems, including ensuring higher variability in the data sets used for training, paying special attention to the labeling that is used with supervised learning algorithms, and considering many-to-one mappings that align with users' perceptions of what makes gestures similar or different in gesture-based interaction systems.

**ACKNOWLEDGMENTS**

The authors would like to thank Eric Gallo and Tim Shea for their comments and suggestions and Taylor Tabb for his assistance in preparing the video preview. We would also like to thank the reviewers for their thoughtful and thorough feedback.**REFERENCES**

[1]  Alexander, J., Han, T., Judd, W., Irani, P. and Subramanian, S. 2012. Putting Your Best Foot Forward: Investigating Real-World Mappings for Foot-based Gestures. *CHI '12 Proceedings of the SIGCHI Conference on Human Factors in Computing Systems*